\documentclass{article}

     \PassOptionsToPackage{square,sort,comma,numbers}{natbib}

     \usepackage[preprint]{neurips_2020}

\usepackage[utf8]{inputenc} 
\usepackage[T1]{fontenc}    
\usepackage{hyperref}       
\usepackage{url}            
\usepackage{booktabs}       
\usepackage{amsfonts}       
\usepackage{nicefrac}       
\usepackage{microtype}      
\usepackage{graphicx}
\usepackage{multirow}
\usepackage{comment}
\usepackage{float}

\title{Fixed-Length Protein Embeddings using Contextual Lenses}

\author{%
  Amir Shanehsazzadeh \\
  Harvard University\\
  Cambridge, MA 02138 \\
  \texttt{amirshanehsazzadeh@college.harvard.edu} \\
  \AND
  David Belanger \\
  Google Research \\
  \texttt{dbelanger@google.com}
  \And
  David Dohan \\
  Google Research \\
  \texttt{ddohan@google.com}
}

\begin{document}

\maketitle

\begin{abstract}
The Basic Local Alignment Search Tool (BLAST)~\cite{Altschul1990} is currently the most popular method for searching databases of biological sequences. BLAST compares sequences via similarity defined by a weighted edit distance, which results in it being computationally expensive. As opposed to working with edit distance, a vector similarity approach can be accelerated substantially using modern hardware or hashing techniques~\cite{avq_2020}. Such an approach would require fixed-length embeddings for biological sequences. There has been recent interest in learning fixed-length protein embeddings using deep learning models under the hypothesis that the hidden layers of supervised or semi-supervised models could produce potentially useful vector embeddings. We consider transformer (BERT~\cite{bert}) protein language models that are pretrained on the TrEMBL data set~\cite{trembl} and learn fixed-length embeddings on top of them with contextual lenses~\cite{lenses}. The embeddings are trained to predict the family a protein belongs to for sequences in the Pfam database~\cite{ElGebali2018, Finn2013}. We show that for nearest-neighbor family classification, pretraining offers a noticeable boost in performance and that the corresponding learned embeddings are competitive with BLAST. Furthermore, we show that the raw transformer embeddings, obtained via static pooling, do not perform well on nearest-neighbor family classification, which suggests that learning embeddings in a supervised manner via contextual lenses may be a compute-efficient alternative to fine-tuning.  
\end{abstract}

\section{Introduction}
Our goal is to assess the viability of replacing BLAST's~\cite{Altschul1990} dynamic programming approach with a vector similarity approach using fixed-length neural network embeddings. These embeddings will have to possess signal for protein features such as evolutionary relationships, structure, or function to be useful. Neural network generative models have become a popular tool for modeling proteins~\cite{Riesselman2018, unirep, bepler, tape2019, bertology, Rives2019, performer}. To apply these models to downstream tasks, embeddings are generated from the intermediate layers, generally with some kind of pooling. The above works have all shown evidence of transfer learning, where pretraining on unlabeled protein sequences results in improved downstream performance. 

Contextual lensing is a general method of producing fixed-length vector embeddings, originally designed for sentences~\cite{lenses}. A contextual lens is a mapping that potentially includes learnable and self-attentive components and that applies some form of reduction, via pooling, over the sequence-length-dependent axis of a model's sequence representation. The motivation for this approach is to avoid model fine-tuning, which is expensive and sensitive to hyperparameter choice, and instead freeze the encoder, which is generally a transformer or BERT~\cite{bert} model, and learn a mapping on top of this encoder's embeddings. In Kiros et. al. ~\cite{lenses}, applying contextual lenses to BERT representations produced embeddings that were competitive with existing state-of-the-art results in NLP, which used models that required fine-tuning. TAPE's~\cite{tape2019} ``attention-weighted mean,'' which was used to learn fixed-length protein embeddings, can also be regarded as a lens.

In this paper, we show that a contextual lens can be used to learn robust embeddings from a pretrained transformer model that provide accurate protein family classification. To learn embeddings we treat the transformer as a frozen encoder and apply a learnable contextual lens. These learned embeddings are trained to predict the families of a group of sequences from the Pfam database that belong to a set of \emph{lens families}. We then use this trained lens, along with the transformer encoder, to predict protein families belonging to a distinct set of families, which we refer to as \emph{evaluation families}, by computing protein embeddings and using a nearest-neighbors classifier (1-NN). We find that this approach is competitive with 1-NN classification using BLAST's weighted edit distance, which suggests potential for a ``Neural BLAST'' that searches for sequences using embedding similarity. Our code is publicly available\footnote{\href{https://github.com/googleinterns/protein-embedding-retrieval}{https://github.com/googleinterns/protein-embedding-retrieval}}.

\section{Neural BLAST}
We represent proteins using an amino acid level encoding. A \emph{lens architecture} is defined to be the combination of an encoder (one-hot encoding, CNN, RNN, transformer),  2-D per-residue representation to a 1-D fixed-length embedding, and a dense predictor layer. For contextual lenses we consider MeanPool, MaxPool, and LinearMaxPool (learnable linear transformation followed by MaxPool). More details can be found in Appendix A.

For training and evaluation we consider the task of Pfam~\cite{ElGebali2018, Finn2013} family classification where we learn a mapping from a protein's primary sequence to its family. We take all sequences belonging to 10000 of the 17129 families, which we denote as \emph{lens families}, and we do the same for 1000 disjoint families, which we denote as \emph{evaluation families}. The lens families are used for \emph{lens training} and the evaluation families are used for evaluating the embeddings with KNN.

\subsection{Lens Training}
Using the train-test split in~\cite{Bileschi2019} for the sequences belonging to the 10000 lens families, we train a lens architecture to directly perform family classification for these families. We measure the accuracy of this trained lens architecture on the test subset of the lens families, which we refer to as \emph{Lens Accuracy}. 

\subsection{Evaluating Protein Embeddings for Nearest-Neighbor Retrieval}
Our objective is to assess the viability of a Neural BLAST that searches sequences via vector similarity on neural network protein embeddings. To see if our model produces embeddings that are effective for such a task, we measure the performance of a KNN (1-NN) classifier on the embeddings of the proteins in the evaluation families. The idea here is that we have trained a model to classify the proteins belonging to the lens families, and we want to see if the embeddings it produces generalize and separate a distinct group of evaluation families.

We thus take the lens architecture, trained using lens training, and use it to compute fixed-length embeddings for the sequences belonging to the evaluation families. Again using the train-test split in~\cite{Bileschi2019}, but this time for the sequences belonging to the 1000 evaluation families, we fit a 1-NN family classifier using the sequence embeddings. We fit using at most $n$ samples per family from the train subset of the evaluation families and consider $n \in \{1, 5, 10, 50\}$ to see how well our embeddings perform with limited data. Finally, we take this 1-NN classifier and measure its accuracy on the test subset of the evaluation families. We refer to this metric as \emph{$n$-Sample Test KNN Accuracy}. As a baseline we measure the accuracy of ``BLAST-KNN'' which is 1-NN using BLAST's weighted edit distance~\cite{Altschul1990}. Note that BLAST-KNN can be thought of as simply returning the family of the top search hit in the database.

\section{Related Work}
Bileschi et al.~\cite{Bileschi2019} develops ProtCNN, a deep, residual, dilated CNN trained for an analogous Pfam family classification task. This model, when trained on $\sim$1.1M samples corresponding to 17129 families using a random train-test split, achieves an error rate of 0.495\% (accuracy of 99.505\%). They ensemble 13 different ProtCNN models using majority voting to create ProtENN, which achieves an error rate of 0.159\% (accuracy of 99.841\%). These CNN models outperform variants of BLAST~\cite{Altschul1990} models, which achieve $\sim$1.14-1.654\% error rate. Senter et al.~\cite{Senter} considers the same high-level question as us of using protein embeddings for a vector similarity-based search. They train neural network models for classifying proteins from the RefSeq~\cite{OLeary2015} database and find that such a training procedure, which is analogous to our lens training, produces embeddings that generalize to and separate unseen classes of proteins, allowing for effective nearest-neighbors classification. For more related works see Appendix B.

\section{Models}
Our lens architectures consist of three components: an encoder, a lens, and a dense predictor layer. The encoders we use are either CNN models or frozen transformer (BERT) models. For the CNN models we baseline a simple 2-layer CNN with kernel sizes 5 and feature sizes 1024 and also compare to the more complicated ProtCNN model~\cite{Bileschi2019}, which is a deep, residual, dilated CNN. By frozen transformer we mean a transformer model that has frozen weights. We experiment with a `Small' 2-layer transformer, a `Medium' 6-layer transformer, and a `Large' 36-layer transformer. We test pretrained versions of the transformer models as well. Pretraining is done on the TrEMBL data set~\cite{trembl}. The pretraining procedure involves truncation to a maximum length of 512 and training using the BERT objective~\cite{bert} with the Adam optimizer~\cite{adam}. The exact procedure is specified in Choromanski et al.~\cite{performer} and their code is publicly available\footnote{\href{https://github.com/google-research/google-research/tree/master/protein_lm}{https://github.com/google-research/google-research/tree/master/protein\_lm}}. Our training procedure is described in Appendix C.


\section{Results}

\begin{table}[h!]
  \caption{Best $n$-Sample Test KNN Accuracy for different model types and $n \in \{1, 5, 10, 50\}.$}
  \label{model_type_results}
  \centering
  \begin{tabular}{ccccc}
    \toprule
     & 1-Sample & 5-Sample  & 10-Sample & 50-Sample\\
     \hline
     \hline
    BLAST-KNN & 0.861 & 0.978 & 0.991 & 0.996\\
    \cmidrule{2-5}
    2-Layer CNN & 0.688 & 0.871 & 0.915 & 0.957 \\
    \cmidrule{2-5}
    Transformer & 0.779 & 0.921 & 0.957 & 0.981\\
    \cmidrule{2-5}
    Pretrained Transformer & \textbf{0.874} & 0.974 & 0.985 & 0.996\\
    \cmidrule{2-5}
    ProtCNN~\cite{Bileschi2019} & \textbf{0.877} & \textbf{0.981} & \textbf{0.992} & 0.994\\
    \bottomrule
  \end{tabular}
\end{table}

\begin{figure}
  \centering
  \includegraphics[width=0.43\linewidth]{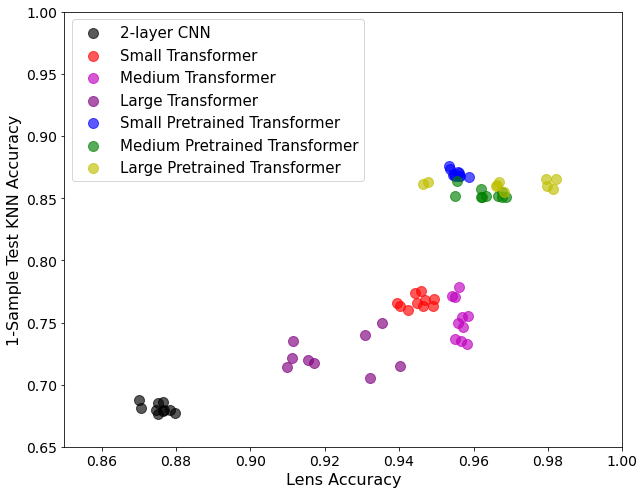}
  \hspace{4mm}
  \includegraphics[width=0.43\linewidth]{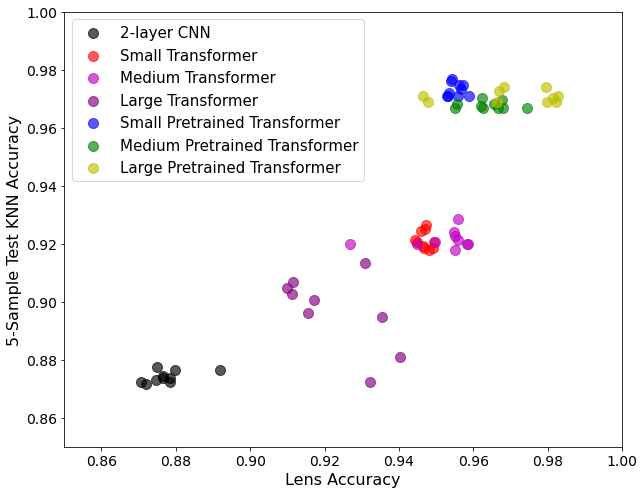}
  \\
  \vspace{3.5mm}
  \includegraphics[width=0.43\linewidth]{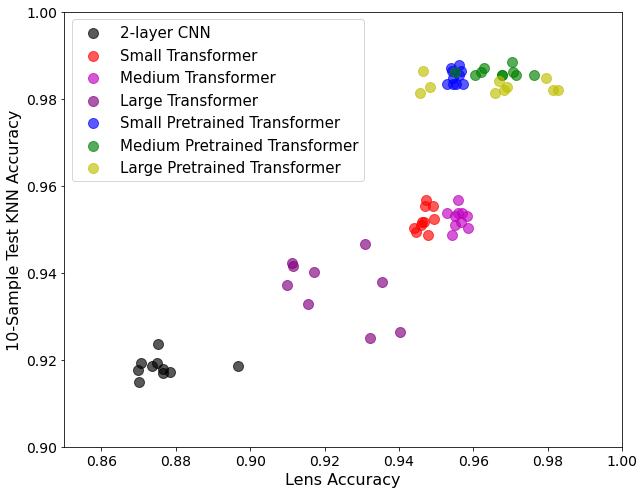}
  \hspace{4mm}
  \includegraphics[width=0.43\linewidth]{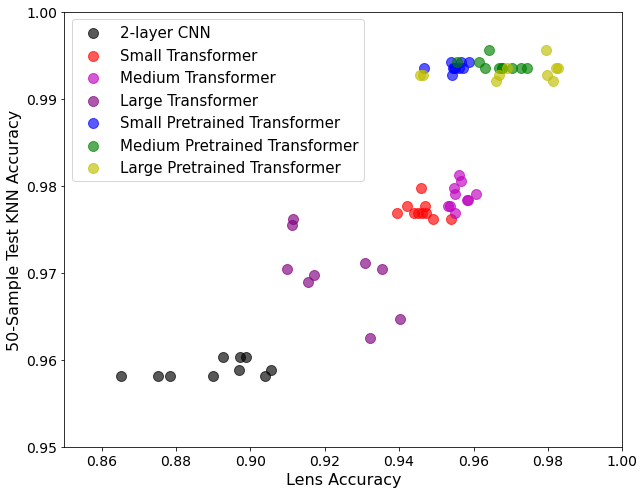}
  \caption{Top 10 $n$-Sample Test KNN Accuracy vs. Lens Accuracy for $n \in \{1, 5, 10, 50\}$. Transformer lens architectures all use the LinearMaxPool lens.}
  \label{test_vs_lens}
\end{figure}

In Table~\ref{model_type_results} we show the best (over random initializations and hyperparameter combinations) $n$-Sample Test KNN Accuracy for different model types. We bold any performance that beats the BLAST-KNN baseline. Figure~\ref{test_vs_lens} shows the top 10 $n$-Sample Test KNN Accuracy measurements vs Lens Accuracy for $n\in\{1, 5, 10, 50\}$ and for different model types. Table~\ref{transformer_results} shows the best KNN accuracy results for all transformer models with all lenses. Recall that the MeanPool and MaxPool lenses are not learnable and so the corresponding embeddings are not directly influenced by the Pfam labels. For the randomly initialized transformer models, we took the maximum performance over 5 random seeds for MaxPool and MeanPool. 


Table~\ref{model_type_results} shows that the pretrained transformer lens architectures outperform BLAST-KNN on 1-sample accuracy, but underperform as the number of samples increases. We also see that the ProtCNN~\cite{Bileschi2019} outperforms the lens architectures for any number of samples and beats BLAST-KNN for 1, 5, and 10 samples. Note that with the LinearMaxPool lens the pretrained transformer lens architectures significantly outperform the randomly initialized transformer lens architectures, especially when the error rate is considered.  Figure~\ref{test_vs_lens} shows that with the LinearMaxPool lens, pretrained transformer lens architectures outperform randomly initialized transformer lens architectures on both Lens Accuracy and $n$-Sample Test KNN Accuracy regardless of transformer size and relatively independently of random initialization or training hyperparameters. Table~\ref{transformer_results} shows that the pretrained transformer + LinearMaxPool lens architecture noticeably outperforms the randomly initialized transformer + LinearMaxpool lens architecture and that the non-learnable lenses considerably underperform.

\setlength{\tabcolsep}{5.625pt}
\begin{table}
  \caption{Best $n$-Sample Test KNN Accuracy for all transformers and $n \in \{1, 5, 10, 50\}.$ MeanPool and MaxPool lenses are static so their embeddings are not directly influenced by Pfam labels.}
  \label{transformer_results}
  \centering
  \begin{tabular}{ccccccc}
    \toprule
     & & & 1-Sample & 5-Sample  & 10-Sample & 50-Sample\\
     \hline
     \hline
     \multirow{9}{*}{Not Pretrained} & \multirow{3}{*}{MeanPool} & Small & 0.398 & 0.639 & 0.719 & 0.830 \\
    &  & Medium & 0.423 & 0.663 & 0.742 & 0.848\\
    & & Large & 0.412 & 0.655 & 0.740 & 0.848\\
        \cmidrule{3-7}
     & \multirow{3}{*}{MaxPool} & Small & 0.427 & 0.704 & 0.786 & 0.898 \\
     & & Medium & 0.517 & 0.790 & 0.854 & 0.932\\
     & & Large & 0.518 & 0.788 & 0.858 & 0.926\\
             \cmidrule{3-7}

     & \multirow{3}{*}{LinearMaxPool} & Small & 0.769 & 0.921 & 0.952 & 0.974 \\
    & & Medium & 0.779 & 0.921 & 0.957 & 0.981\\
    & & Large & 0.750 & 0.895 & 0.938 & 0.970\\
    \cmidrule{2-7}
    \cmidrule{2-7}
    \multirow{9}{*}{Pretrained} & \multirow{3}{*}{MeanPool} & Small & 0.359 & 0.634 & 0.721 & 0.841 \\
    & & Medium & 0.399 & 0.668 & 0.752 & 0.859\\
    & & Large & 0.367 & 0.634 & 0.717 & 0.824\\
        \cmidrule{3-7}
    & \multirow{3}{*}{MaxPool} & Small & 0.419 & 0.683 & 0.770 & 0.881 \\
    & & Medium & 0.395 & 0.647 & 0.739 & 0.864\\
    & & Large & 0.389 & 0.656 & 0.736 & 0.851\\
    \cmidrule{3-7}
    & \multirow{3}{*}{LinearMaxPool} & Small & \textbf{0.874} & 0.969 & 0.980 & 0.993 \\
    & & Medium & \textbf{0.864} & 0.968 & 0.985 & 0.994\\
    & & Large & \textbf{0.866} & 0.974 & 0.985 & 0.994\\
    \bottomrule
  \end{tabular}
\end{table}

\section{Discussion}
It is encouraging to see that neural networks, in particular pretrained transformer models and ProtCNN~\cite{Bileschi2019}, can outperform the BLAST-KNN. This suggests that a Neural BLAST database search with comparable performance to BLAST but faster query time may be feasible. The results in Table~\ref{model_type_results} and Table~\ref{transformer_results} show some degree of transfer learning from the model pretraining as the pretrained transformers coupled with the LinearMaxPool lens outperform their non-pretrained counterparts regardless of model size or number of samples. Figure~\ref{test_vs_lens} not only supports this but also suggests that pretrainin offers an improvement in the distribution of performances. In Table~\ref{transformer_results} we see that both pretrained and non-pretrained models perform poorly with the static MeanPool and MaxPool lenses compared to the performance with the learnable LinearMaxPool lens. This indicates generalization from the lens training task and the corresponding performance boost is comparable to model fine-tuning, but requires substantially less compute.

Our results indicate that contextual lenses~\cite{lenses} offer a relatively inexpensive approach for the supervised learning of fixed-length protein embeddings that are useful for family classification. Additionally, we show that while the raw pretrained transformer embeddings, obtained via applying MaxPool or MeanPool, do not themselves perform well on nearest-neighbor family classification, the learnable LinearMaxPool lens is able to bring out the increased signal from the model pretraining to create more robust embeddings. This motivates us to suggest that contextual lenses be considered as an efficient alternative to model fine-tuning for downstream applications. We speculate that this framework can be used to learn embeddings designed for other downstream tasks, such as landscape or secondary structure prediction, and that these embeddings can be used to create a (potentially task-specific) Neural BLAST that performs sequence search via embedding similarity.

\bibliographystyle{plain}
\bibliography{main}

\section*{Appendix}

\subsection*{A \ \ \ \ Background}

\subsubsection*{A.1 \ \ \ \   Proteins}
We consider proteins using only their primary structure, that is their amino acid sequence, with a 21-letter alphabet that includes the 20 standard amino acids~\cite{1984} as well as a pad index. A length $\ell$ protein $a = a_1a_2\cdots a_\ell$ is thus modeled as a discrete sequence $x = (x_1, x_2, ..., x_\ell)$ with $x_i \in \{0, 1, ..., 19\}$ and potentially padded to $(x_1, x_2, ..., x_{\ell}, 20, 20, ..., 20)$.

\subsubsection*{A.2 \ \ \ \ Model Embeddings and Contextual Lenses}
In general, an encoder model $\Phi$ (one-hot encoding, CNN, RNN, transformer) maps a protein $x$ of length $\ell$ to a sequence-length-dependent array representation: $\Phi(x) \in \mathbb R^{n \times \ell}$ for some constant $n$. For learning fixed-length embeddings the goal is to learn another mapping $\Psi$ such that $z=\Psi(\Phi(x)) \in \mathbb R^m$ with $m$ being independent of $\ell$ and $z$ being a useful protein embedding, in the sense of having strong signal for some downstream feature.

The simplest contextual lens~\cite{lenses} is pooling. This lens is non-learnable and involves either taking the average  (MeanPool) or the maximum (MaxPool) over the length-dependent axis of $\Phi(x) \in \mathbb R^{n\times\ell}$. A learnable, but simple, contextual lens involves a single dense layer with weight matrix $W \in \mathbb R^{m \times n}$ and bias vector $b \in \mathbb R^m$ and a non-linear activation $\phi$. Then we consider the array of transformed amino acid level embeddings $$\left[\phi(W\Phi(x)^{(i)}+b)\right]_{i=1}^{\ell} \in \mathbb R^{m \times \ell}$$ and apply a pooling operation. If we apply MaxPool we call this operation LinearMaxPool and likewise for MeanPool. Note that we only use the ReLU activation for $\phi$. A contextual lens that is both learnable and self-attentive involves a form of gated convolution. However, since we do not use this lens, we defer discussion of it to the original paper.

\subsection*{B \ \ \ \ Additional Related Work}
We are unaware of other directly comparable works that attempt the same task, but machine learning techniques, specifically language models, have been used in a similar fashion to learn useful protein embeddings in~\cite{Riesselman2018, unirep, bepler, tape2019, Rives2019, bertology}.

TAPE~\cite{tape2019} proposes 5 downstream benchmark tasks and baselines a number of language models. Their language models, when pretrained on the Pfam database~\cite{ElGebali2018, Finn2013}, perform significantly better. However, they did not find that performance on the language modeling task correlates strongly with downstream performance, which is a key motivating result from NLP. 

RNN protein language models were developed by Alley et al.~\cite{unirep} and by Bepler et al.~\cite{bepler}.  UniRep is a unidirectional multiplicative LSTM protein language model trained on UniRef50~\cite{Suzek2014}. UniRep learns embeddings that encode relevant protein information such as amino acid biochemistry, secondary structure, and evolutionary and functional information. Bepler et al. train a bidirectional LSTM that they pretrain on Pfam sequences~\cite{ElGebali2018, Finn2013}. The learned embeddings are compared using a soft symmetric alignment and are useful for structure classification and transmembrane prediction.

Rives et al.~\cite{Rives2019} does similar work but trains varying size transformer language models on 250 million UniRef~\cite{Suzek2014} sequences. Their transformer embeddings also encode amino acid biochemistry and achieve state-of-the-art performance on a number of downstream tasks such as contact,  secondary structure, and remote homology. Additionally, they show that larger model size, which results in better protein language modeling performance, improves downstream performance. Vig et al.~\cite{bertology} consider transformer models as well and show that the attention mechanism itself, computed from pretrained protein BERT models, learns relevant features such as distance in 3-D, location of binding sites, and contact maps.

\subsection*{C \ \ \ \ Training Procedure}
For lens training, we use the Adam optimizer~\cite{adam} with variable learning rates and weight decays per lens architecture component, but with no learning rate warmup. Cross-entropy loss is used. Sequences are padded or truncated to a maximum length of 512 and padded components are zeroed out before pooling. We do a grid search over learning rates, weight decays, and random initialization seeds. For the transformer lens architectures the encoder learning rate and weight decays are always set to 0 since the transformer is frozen. Training hyperparameter combinations can be found in our GitHub repo\footnote{\href{https://github.com/googleinterns/protein-embedding-retrieval/blob/master/params_combinations.json}{https://github.com/googleinterns/protein-embedding-retrieval/blob/master/params\_combinations.json}}. For KNN classifiers we use scikit-learn's~\cite{scikit-learn} KNN implementation.

\end{document}